# Interfacial exchange and magnetostatic coupling in a CoFeB/Thulium Iron Garnet heterostructure

Walid Al Misba [1], Jenae E. Shoup [2], Miela J. Gross [3,4], Dhritiman Bhattacharya [5], Kai Liu [5], Caroline A. Ross [4], Daniel B. Gopman [2], Jayasimha Atulasimha [1,6,7*]

[1] Mechanical and Nuclear Engineering, Virginia Commonwealth University, Richmond, VA, USA
[2] Materials Science & Engineering Division, National Institute of Standards and Technology, Gaithersburg, MD, USA
[3] Electrical Engineering and Computer Science, Massachusetts Institute of Technology, Cambridge, MA, USA
[4] Department of Materials Science and Engineering, Massachusetts Institute of Technology, Cambridge, MA, USA
[5] Department of Physics, Georgetown University, Washington, D.C, USA
[6] Electrical and Computer Engineering, Virginia Commonwealth University, Richmond, VA, USA
[7] Department of Physics, Virginia Commonwealth University, Richmond, VA, USA
*jatulasimha@vcu.edu

**Abstract:**

We investigate the exchange coupling between a ferrimagnetic insulator (FI) thulium iron garnet (TmIG) deposited on a gadolinium gallium garnet (GGG) substrate, which shows perpendicular magnetic anisotropy and a ferromagnetic metal (FMM) stack with oxide capping that consists of CoFeB($x$)/W(0.4 nm)/CoFeB(0.8 nm)/MgO(1 nm)/W(5 nm). Vibrating sample magnetometry, magneto-optical Kerr microscopy and first-order reversal curve studies coupled with micromagnetic simulations are used to analyze the coupling between these layers. Strong interlayer exchange coupling and magnetostatic coupling are observed in the samples where the relative strength between these interactions can be controlled by varying the thickness of the CoFeB layer. Films with CoFeB thickness $x \leq 1$ nm are strongly exchange coupled, whereas the magnetostatic coupling dominates when the thickness is increased to 3 nm or above. These findings have important implications towards realizing fast and energy efficient spintronic devices using a FI, as its coupling to the FMM layer can be used for effective electrical read out of the magnetic state of the FI.

## I. Introduction:

Rare-earth iron garnets (REIG, $RE_3Fe_5O_{12}$, RE=thulium, terbium, samarium, etc.) with perpendicular magnetic anisotropy (PMA) have been developed for spintronic applications [1,2]. Heterostructures such as Pt/TmIG (thulium iron garnet) and Pt/BiYIG (Bi-substituted yttrium iron garnet) have demonstrated spin orbit torque switching [1,3,4], chiral spin textures [3,5,6], and a relativistic domain wall velocity approaching the magnon group velocity [7], making these materials promising for computing devices. REIG materials exhibit ferrimagnetic characteristics due to the antiparallel superexchange coupling between the tetrahedral and octahedral Fe sublattices [8]. However, the insulating nature of the REIG precludes its incorporation into a magnetic tunnel junction (MTJ). Coupling the REIG to a ferromagnetic metal (FMM) would enable readout of the magnetization of the REIG, as the FMM can form the free layer of an MTJ. FMM/REIG multilayers could also be useful in magnon-based spintronics for high frequency application and low power computing. This motivates study of the strength and mechanism of exchange coupling between thin films of magnetic metals and REIGs.

The ground state of a magnetic heterostructure is determined through the competition between long-range magnetostatic interactions, short-range exchange interactions, and local anisotropy variations that are also determined by the composition and thickness of the constituent layers. Previously, exchange-coupled ferromagnetic multilayers in direct contact [9-11] and separated by a non-magnetic spacer [12-16] and hybrid spacer [17] were studied. Dynamic coupling between the ferrimagnetic insulator (FI) and FMM due to exchange of non-equilibrium spin currents were identified by investigating broadband ferromagnetic resonance and spin torque ferromagnetic resonance [18-20]. However, few studies have been performed on the static coupling in FI/FMM heterostructures [21-24]. Sputtered Fe films of 5 nm to 10 nm thickness grown above 100 nm thick YIG were shown to replicate the stripe domain wall pattern of YIG due to magnetic coupling [22]. In addition, static and dynamic magnetization behavior has been studied in heterostructures combining 0.5 μm thick $(YBiLu)_3(FeAl)_5O_{12}$ with a 30 nm thick permalloy film [23]. In that system, strong exchange coupling led to the increased domain wall periodicity after permalloy deposition and no domain wall imprinting was reported. In another study, a 2 nm Co thin film was deposited on YIG films of varying thickness in the μm range, and changes in domain wall geometry of the YIG films was attributed to magnetostatic coupling [24]. In summary, while past work has suggested coupling between FMMs and FIs, the exact mechanism (exchange vs. magnetostatic), their relative importance as a function of the FMM layer thickness, and an estimate of the exchange coupling strength has not been studied.

In this study, we examine thulium iron garnet ($Tm_3Fe_5O_{12}$, or TmIG) as the FI layer. When deposited on a (111) gadolinium gallium garnet (GGG) substrate, TmIG is under in-plane tension and exhibits a PMA of magnetoelastic origin. To fabricate the FI/FMM heterostructure, an FMM stack consisting of CoFeB($x$)/W(0.4 nm)/CoFeB(0.8 nm)/MgO(1 nm)/W(5 nm) as shown in Fig. 1a is deposited on top of the FI using ion beam sputtering. The thickness of CoFeB in direct contact with TmIG is varied to investigate the extent of exchange and magnetostatic interaction, while the W acts as a diffusion barrier for the boron from CoFeB and the 1 nm MgO layer deposited on top simulates the role of the insulating tunnel barrier layer of a MTJ device. Advantageously, the MgO also promotes PMA in the CoFeB layer below, reinforcing exchange and magnetostatic coupling from the PMA FI on the FMM stack. This artificial layered structure was chosen to combine low coercivity and PMA in the TmIG with the ability to incorporate CoFeB as the soft layer of CoFeB/MgO/CoFeB-based MTJs.

Figure 1 depicts the heterostructures investigated, where the FI is perpendicularly magnetized while the magnetization of the FMM stack can be oriented out of plane, within the plane, or some admixture of the two, depending on the thickness of the CoFeB overlayer. Vibrating sample magnetometry (VSM) and magnetic imaging with magneto-optical Kerr microscopy (MOKE) is performed to investigate the coupling phenomena. First-order reversal curves (FORCs) are obtained to analyze the magnetic interactions in the heterostructure, and micromagnetic simulations based on experimental behavior are performed to quantify the extent of competing interactions from the interlayer exchange and magnetostatic interaction.

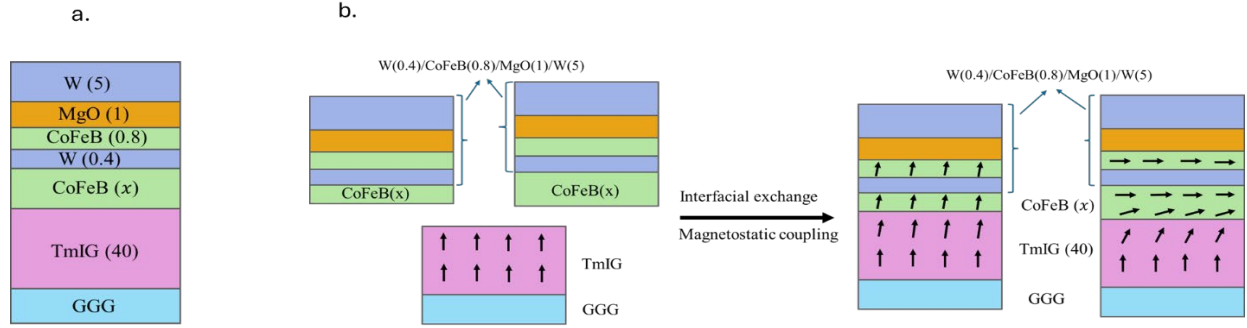

**Figure 1: a**. The prepared FMM stack CoFeB($x$)/W(0.4 nm)/CoFeB(0.8 nm)/MgO(1 nm)/W(5 nm) are deposited on FI stack GGG/TmIG(40 nm) to investigate FI/FMM coupling. All the numbers in the parentheses are thickness in nm. **b.** The FMM stacks with CoFeB of $x = 1$ nm and lower thickness have magnetization predominantly oriented along the out of plane direction whereas in stacks with $x = 3$ nm and higher thickness the magnetizations are predominantly along the in-plane. The resulting magnetizations in FI/FMM stacks are canted with respect to out of plane and in-plane due to interfacial exchange coupling and magnetostatic interactions.

## II. Sample Preparation

A 40 nm thick TmIG was deposited on top of the GGG (111) substrate using a pulsed laser deposition (PLD) system with base pressure $\leq 7 \times 10^{-4}$ Pa ($5 \times 10^{-6}$ Torr). After loading the substrates, a 20 Pa (150 mTorr) $O_2$ partial pressure environment is maintained through continuous $O_2$ flow during substrate temperature ramp-up to 650 °C, during the deposition, and while cooling down post-deposition. The 248 nm KrF excimer laser was focused to a fluence of 2 J/cm² on a stoichiometric TmIG target at a frequency of 10 Hz.

The GGG/TmIG samples are transferred to a 12-target, 200 mm ion beam sputtering cluster with base pressure $\leq 3 \times 10^{-6}$ Pa ($2 \times 10^{-8}$ Torr). In this chamber, thin films of CoFeB($x$)/W(0.4 nm)/CoFeB(0.8 nm)/MgO(1 nm)/W(5 nm) are grown in an Ar working pressure of 0.03 Pa ($2 \times 10^{-4}$ Torr). Using a $Co_{20}Fe_{60}B_{20}$ stoichiometric target, a thickness series where $x$ ranges across the nominal thicknesses: 0.8 nm, 1 nm, 3 nm and 4 nm is deposited. A control sample for each CoFeB thickness was deposited on a Si substrate with a 300 nm thick thermally oxidized Si overlayer simultaneously with the TmIG/GGG samples. All samples were processed in a rapid thermal annealer under vacuum at 300 °C for 10 min. The W spacer layer is employed as a boron sink during the annealing and to achieve PMA in the CoFeB layer [25].

## III. Sample Characterization:

The control samples are characterized by measuring the hysteresis loops for in-plane and out of plane external fields using VSM as shown in Fig. 2. The thickness of different layers of the heterostructure and corresponding saturation magnetization, easy axis direction, and the coercive field along the easy axis are presented in Table SI in Supplemental Information. When the thickness of the lower CoFeB layer is $x = 1$ nm or less, FMM stack exhibits PMA due to the W and exchange coupling from the top PMA CoFeB layer. For $x = 3$ nm and above, the FMM stack shows a predominantly in-plane easy axis attributed to dominant magnetostatic effects.

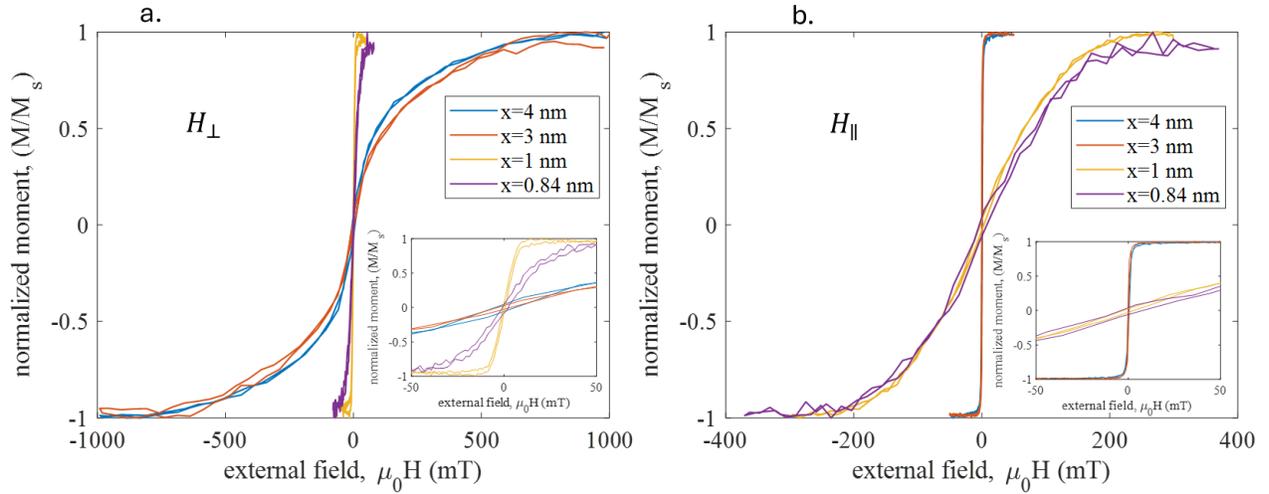

**Figure 2: a.** Out of plane and **b.** in-plane hysteresis loops for control samples, Si/CoFeB(x)/W(0.4 nm)/CoFeB(0.8 nm)/MgO(1 nm)/W(5 nm) for variable thickness, $x$ of CoFeB.

Furthermore, there is no appearance of multiple steps or zero moment crossing in the hysteresis loops, which suggests substantial ferromagnetic coupling between the two CoFeB layers above and below the dusting W layer. Thus, the fixed 0.8 nm CoFeB layer above W, which is in contact with a top MgO and expected to have PMA [26], remains PMA for $x \leq 1$ nm thick CoFeB layer, while it can become predominantly in-plane when coupled with the $x \geq 3$ nm thick CoFeB.

Next, the magnetic hysteresis loops of the FI/FMM heterostructures are obtained for variable CoFeB thickness using VSM and shown in Fig. 3a-3d. The thickness of different layers of the heterostructure of the FI-only stack and the FI/FMM stacks along with the measured saturation magnetization, easy axis directions, and the coercive field along the easy axis directions are presented in Table SII in the supplement. Two distinct trends are seen between the heterostructure samples with CoFeB thickness of $x \leq 1$ nm (Fig. 3a-3b) and $x \geq 3$ nm (Fig. 3c-3d). For comparison, the hysteresis loops of pristine GGG/TmIG (FI-only stack) are also presented. The out of plane coercivity of the FI/FMM increases to $1.81 \pm 0.1$ mT for low thickness CoFeB ($x = 1$ nm), compared to the coercivity of the pristine FI of $0.22 \pm 0.1$ mT and control FMM stack which is $0.52 \pm 0.1$ mT for $x = 1$ nm. The out of plane hysteresis loop of the FI/FMM has no steps, suggestive of magnetostatic and/or exchange coupling between the FI and FMM that causes them to switch together. In addition, the FI/FMM samples show a more gradual transition towards saturation in Fig. 3a. The exchange coupling also modifies the PMA of FI/FMM compared to individual FI and FMM stacks. The PMA modification is evident from the in-plane hysteresis loops in Fig. 3b which shows the FI/FMM samples saturate at higher fields, $\mu_0 H_\parallel > 130$ mT (increase in the hard axis anisotropy) compared to the FI-only sample which saturates at 100 mT.

In thick CoFeB heterostructures, with $x \geq 3$ nm, the hysteresis loops in Fig. 3c-3d are modified with respect to individual FI and FMM loops which again suggests coupling. In contrast to the thin CoFeB heterostructures in Fig. 3a, the out of plane loops in Fig. 3c become more canted and the heterostructure magnetizations are seen to switch in a reversible manner, possibly due to an increased portion of the CoFeB film that tends to be in-plane due to shape anisotropy. The saturation fields, $\mu_0 H_\perp > 900$ mT, show similar values to that of the FMM-stack comprising thick CoFeB layers (Fig. 2a). Furthermore, the in-plane loops of FI/FMM show higher remanent magnetization than the out of plane loops. For instance, the in-plane remanence of FI/FMM with $x = 4$ nm CoFeB, where the TmIG and CoFeB magnetization comprises approximately half of the total magnetization, $M_s$ (Table SII) is found to be $0.56 M_s$ compared to $0.07 M_s$ along the out of plane direction. This suggests the TmIG magnetization at the interface is forced to the canted state due to a combined effect of magnetostatic and exchange coupling, with larger projections along the in-plane CoFeB magnetization. Also, the in-plane coercive fields of FI/FMM are increased compared to the FMM-stack due to the coupling (Table SI and Table SII). The increase is lower in FI/FMM with $x =$ 4 nm CoFeB, as the thicker CoFeB favors a net in-plane magnetization at the TmIG/CoFeB interface.

Each of the studied FI/FMM heterostructure samples yields a single hysteresis loop, which are different than the respective FI and FMM loops. This is attributed to both the exchange interactions and magnetostatic coupling between the magnetic layers. The relative strength between these interactions is further investigated with MOKE microscopy and FORC analyses.

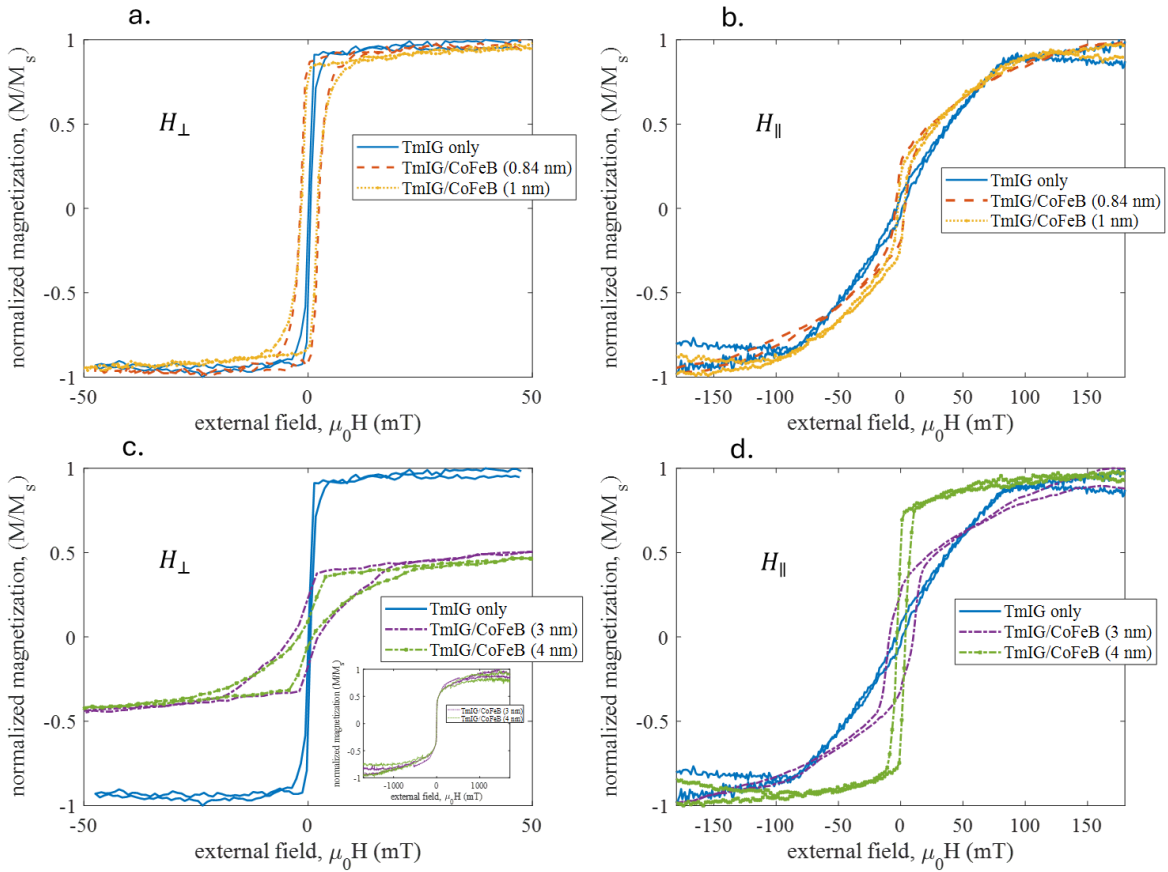

**Figure 3**: **a.** Out of plane and **b.** in-plane hysteresis loops of FI/FMM heterostructure samples, GGG/TmIG(40 nm)/CoFeB(x)/W(0.4 nm)/CoFeB(0.8 nm)/MgO(1 nm)/W(5 nm), with CoFeB thickness, $x \leq 1$ nm. **c.** out of plane

and **d.** in-plane hysteresis loops with CoFeB thickness $x \geq 3$ nm. The out of plane and in-plane hysteresis loops of the pristine GGG/TmIG(40 nm) sample are presented for comparison.

.

## IV. Interlayer exchange and Magnetostatic coupling:

Polar MOKE is carried out to observe the domain pattern for the FI/FMM heterostructure samples during the switching process. The representative MOKE images of the heterostructure and control FMM samples are shown in Fig. 4, for external fields applied along the out of plane direction. The samples are subjected to a positive 50 mT external field and the images are taken by reducing the fields from +50 mT. Control FMM samples with $x \leq 1$ nm CoFeB, as shown in Fig. 4a, exhibit changes only in intensity during switching; however, no signs of domain wall nucleation and propagation are evident. In contrast, clearly distinguishable and regularly spaced labyrinth domain patterns are observed for the TmIG sample (FI-only stack) as in Fig. 4b. The black contrast domains nucleate around -1 mT and expand further with the decreasing field, saturating around -10 mT. To extract the domain wall periodicity, the radial fast fourier transform (FFT) intensity profile of the images is measured as shown in Fig. 4b where the peak location of the Gaussian FFT intensity denotes the predominant domain wall periodicity which is the mean domain wall periodicity (as the distribution is almost symmetric, i.e., Gaussian). The domain wall periodicity of the FI-only sample is estimated to be 4.90 μm. Similar to the FI-only sample but in contrast to the control FMM, the FI/FMM heterostructure with x=1 nm is observed to switch via domain wall nucleation and propagation, as illustrated in Fig. 4c. This indicates a strong imprinting of the magnetic domain structure of TmIG on the CoFeB through exchange coupling. In this sample, significant domain wall nucleation is not observed up to -2 mT which suggests the increased coercivity stemmed from strong interlayer exchange coupling and was further enhanced by the pinning of TmIG magnetization to the CoFeB magnetization. Around -3 mT, dendritic domain patterns with black contrast are observed, which nearly exceed the area of the regions of the sample with white domains. The periodicity of the domain wall reduces to 3.2 μm, which suggests increased resultant magnetic moment of the sample compared to the TmIG sample. The FI/FMM with $x = 0.8$ nm CoFeB behaves similarly to that of the $x = 1$ nm CoFeB sample.

The periodicity of the dendritic domain decreases further with the increase of the CoFeB layer thickness. For FI/FMM with $x = 4$ nm, the domain wall starts to nucleate well before 0 mT suggesting strong magnetostatic interaction accompanied by a large increase in the resulting magnetic moment of the heterostructure. At 0 mT, domain walls with multiple periodicities are observed as seen from the mostly flattened gaussian distribution with one barely distinguishable peak around 0.93 μm. Although the reversed black contrast domains are seen to nucleate early for the thicker samples ($x \geq 3$nm), the domains expand gradually, and a large field is needed to saturate the film, which suggests increasing contribution of shape anisotropy of the CoFeB on the TmIG magnetization. Contrary to the FI/FMM withthin CoFeB, in thicker CoFeB heterostructures tiny sized domains with white contrast can still be seen in the images at -13 mT. This is enabled by a large in-plane component of the CoFeB magnetization, which facilitates flux closure of anti-parallel TmIG domains as shown in Fig. 5b.

The domain wall periodicity values serve as an important indicator of the competition between exchange interaction and magnetostatic coupling. Increased magnetostatic coupling favors a greater number of domain walls and oppositely oriented lateral regions to minimize the dipolar energy. As a result, the domain wall periodicity decreases. The stripe domain width ($L_s$), within the limit when the film thickness is much smaller than the domain period, can be found from the analytical model reported by Kaplan and Gehring [28], $L_s = t e^{(\frac{\pi a}{2}+1)} e^{\frac{\pi \sigma_w}{\mu_0 M_s^2 t}}$, where $t$ is the film thickness, $M_s$ is the saturation magnetization, $\sigma_w$ is the domain wall energy, and $a$ is the model dependent parameter with a value of -0.666. As film thickness and saturation magnetization increases, the model predicts a lower value for the domain wall periodicity.

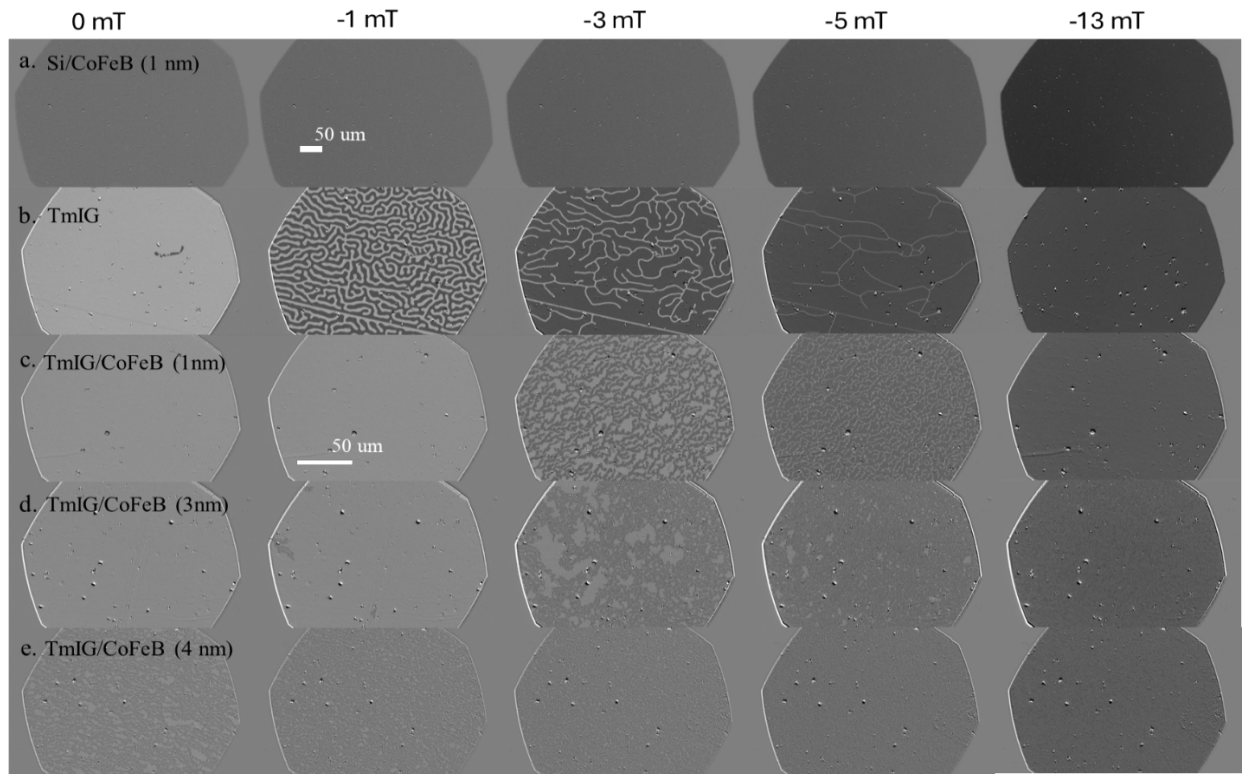

**Figure 4**: Polar MOKE magnetometry for a. FMM stack with $x$ = 1 nm b. FI stack and FI/FMM heterostructure samples with c. $x$ = 1 nm d. $x$ = 3 nm and e. $x$ = 4 nm respectively. The scale bar is the same for b, c, d and e. All the samples are first subjected to +50 mT and then the out of plane external fields are decreased. The snapshots are taken at field values of 0 mT, -1 mT, -3 mT, -5 mT and -13 mT. Labyrinthine domains with regular periodicity are observed for the FI stack. Labyrinthine domains with dendritic patterns are observed with the insertion of CoFeB overlayers in FI/FMM. The periodicity of the domain wall decreases with increased CoFeB thickness implying higher magnetostatic coupling.

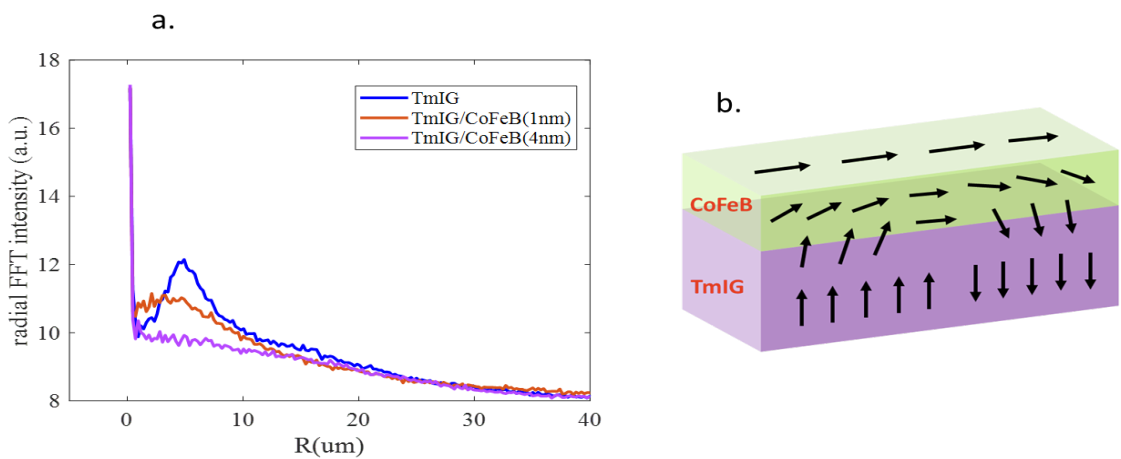

**Figure 5**: a. Radial FFT intensity to determine the periodicity of the domain wall for the TmIG, FI/FMM heterostructures with 1 nm and 4 nm thick CoFeB layers at fields, $\mu_0 H_\perp$ = -1 mT, -3 mT and 0 mT respectively. b. Schematic showing flux closure in FI/FMM heterostructure with CoFeB thickness, $x$ = 3 nm or $x$ = 4 nm.

In the heterostructure samples we studied, the domain wall periodicity decreases with the increase of CoFeB layer thickness, which suggests increasing contributions from magnetostatic coupling. Thus, the domain walls of TmIG imprinted on the PMA CoFeB of thickness $x \leq 1$ nm indicate strong exchange coupling. However, for CoFeB with thickness $x \geq 3$ nm, the TmIG domain patterns do not imprint themselves on the CoFeB through exchange coupling but influence the domain formation that provides flux closure loops.

**First order reversal curves (FORCs)**

We carried out FORCs experiments [29-34] to investigate the interlayer exchange and magnetostatic interactions between the FI and FMM layers. VSM is used to measure the FORCs in the following manner. After driving the sample to positive saturation, the magnetic field is adjusted to a reversal field $H_r$. The magnetization is then measured as the magnetic field, $H$ is gradually increased from $H_r$ until the sample returns to positive saturation, forming a FORC. A series of FORCs is obtained by changing the $H_r$ values in steps, as illustrated in Fig. 6f. The FORC distribution is defined by a mixed second-order derivative:

$\sigma(H_r, H) = \frac{\partial_M^2}{2\partial_{H_r}\partial_H}$ .

Fig. 6c shows the FORC distribution $\sigma$ corresponding to the *M-H* loops in Fig. 6f for the TmIG/CoFeB ($x$ = 1 nm)/W/CoFeB/MgO/W sample subjected to out of plane external fields. Three horizontal line-scans corresponding to the reference circles in Fig. 6f (indicating different reversal fields, $H_r$) are marked in Fig. 6c with dashed lines. For $\mu_0 H_r > 0$ mT, the sample remains mostly saturated, thus, no appreciable FORC features are observed. As $\mu_0 H_r$ is decreased, -3.2 mT $< \mu_0 H_r <$ 0 mT, the magnetization starts to switch by the nucleation and propagation of reversed domains. In this range, a horizontal FORC feature appears in Fig. 6c as denoted by region 1. With further decrease in the reversal field, -10.4 mT $< \mu_0 H_r <$ -3.2 mT, a vertical FORC feature is observed denoted by region 2. The vertical feature in the FORC diagram represents nucleation (annihilation) of positively oriented (reversed) domains after large portions of the samples become negatively saturated. For $\mu_0 H_r < -11$ mT, the sample remains negatively saturated and the successive FORC scans do not show any appreciable change in the magnetization reversal process.

The FORCs distribution for *out of plane fields* of the FI-only (TmIG-only) sample, a representative control FMM stack with $x$ = 1 nm, and FI/FMM with x= 1 nm, 3 nm and 4 nm are presented in Fig. 6a-6e. The FORC diagram of the FI-only sample reveals two features: a horizontal ridge parallel to the $H$ axis and a vertical ridge parallel to the $H_r$ axis, as previously described and typically observed in magnetic structures due to domain nucleation and propagation and subsequent annihilation of domains [30,35-37]. The $x$ = 1 nm CoFeB on Si does not show any FORC features, indicating a Stoner-Wolfarth like reversible behavior with negligible domain wall nucleation as confirmed by MOKE. However, the FI/FMM heterostructure exhibits FORC features again, reinforcing the domain wall nucleation in the MOKE and clearly demonstrating a strong imprinting of the FI texture on the FMM stack for the 1 nm thickness CoFeB layer. Also, in FI/FMM with $x$=1 nm, the horizontal FORC feature shifts to negative values of $H_r$, $\mu_0 H_r$ = -1.2 mT and the vertical ridge shifts to positive values of $H$, $\mu_0 H$ = 1.8 mT compared to the FI-only sample, reinforcing the increased coercivity of the heterostructure [38] as also seen in the VSM curves in Fig. 3. This is attributed to the pinning of the TmIG magnetization from the thin CoFeB layer predominantly due to interlayer exchange coupling. Furthermore, no reversible FORC features around the $H = H_r$ line is observed in Fig. 6c indicating that CoFeB and TmIG magnetizations are switched together with domain wall nucleation and propagation. Similar FORC features are also observed for FI/FMM samples with $x$ = 0.84 nm (not shown), suggesting strong interlayer exchange coupling in samples with $x \leq 1$ nm.

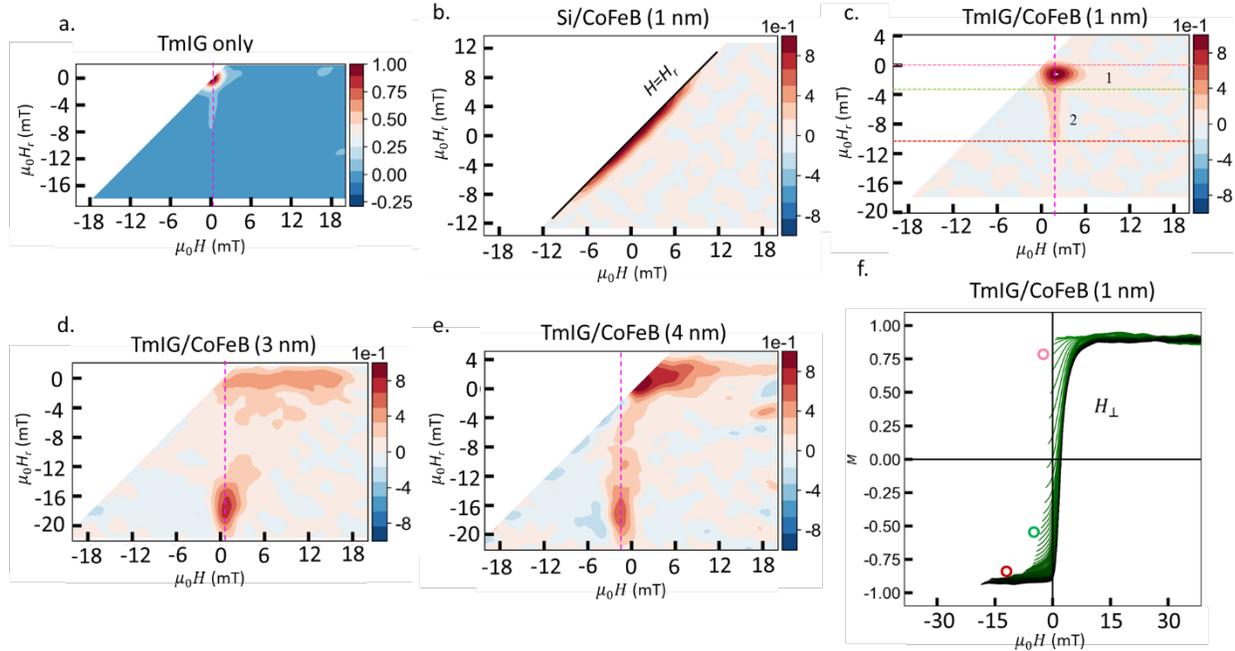

**Figure 6**: FORC distributions for samples with a. FI-only b. control FMM stack with $x$= 1 nm CoFeB and FI/FMM heterostructures with variable CoFeB layer thickness of c. $x$ = 1 nm d. $x$ = 3 nm e. $x$ = 4 nm for out of plane external fields, $H_\perp$ . f. Family of FORC curves determined for different reversal fields, $H_R$ for FI/FMM sample with $x$= 1 nm CoFeB.

Although the FI and FMM layers are coupled and undergo magnetization reversal together within the heterostructure samples, the strength of exchange coupling and magnetostatic interaction varies with the thickness of the CoFeB layer. For instance, in a thick CoFeB heterostructure, such as $x$ = 4 nm, the vertical ridge shifts to negative values of $H$, $\mu_0 H$ = -1.8 mT (Fig. 6e). This indicates that the domain nucleation occurs sooner after saturation, as the in-plane anisotropy of the thick CoFeB layer promotes out of plane switching of the FI/FMM heterostructure at lower fields. This is consistent with increased dipolar interactions, which will promote early domain nucleation [39] during reversal.

The FORC features in the heterostructure with $x$ = 3 nm CoFeB is similar to the heterostructure with $x$= 4 nm CoFeB, except that the vertical ridge shifts to the positive value of $H$, $\mu_0 H$ = 0.6 mT (Fig. 6d). However, the nucleation field is much smaller than the heterostructure sample with x = 1 nm CoFeB. This suggests the dominating interaction in this case is magnetostatic interaction.

Thus, the FORC study corroborates the findings from VSM and MOKE microscopy that the FI and FMM layers are coupled and switch together in the FI/FMM heterostructure. Furthermore, it confirms that interaction is dominated by exchange coupling for samples with $x \leq 1$ nm CoFeB, whereas magnetostatic coupling dominates in samples with $x \geq 3$ nm CoFeB.

## V. Micromagnetic simulation:

Micromagnetic simulations are carried out using Mumax3 [40] to investigate the strength of interlayer exchange interaction for varying CoFeB overlayer thickness in the FI/FMM heterostructures. We choose micromagnetic simulation over the macroscopic Stoner-Wohlfarth model due to its ability to account for non-uniform magnetization, domain nucleation, propagation, and complex magnetic interactions in magnetic multilayers. Excellent qualitative match is found to the hysteresis loops from VSM, and the domain wall patterns obtained from MOKE experiments. A representative thin and thick CoFeB overlayer sample with x = 1 nm and 4 nm respectively are studied. The details of the micromagnetic simulation can be found in the supplementary section S2 (including cell size, parameters used, etc., that are needed to reproduce these simulations). The interlayer exchange coupling (IEC) between TmIG/CoFeB is varied and a qualitative match is found with IEC= 0.25 mJ/m$^2$ and 1.34 mJ/m$^2$ for the $x$ =1 nm and $x$= 4 nm samples respectively.

The main micromagnetic simulation results are presented in Fig. 7 while the supplementary section S3 explains why the above IECs are optimal by showing that higher or lower IECs lead to simulations that are vastly different from experimentally observed behavior.

In the heterostructure with a thin CoFeB sample ($x$ = 1 nm) the domains in TmIG are strongly imprinted on the CoFeB layer as shown in Fig. 7. Also, the top CoFeB layer magnetization exhibits large out of plane projections. Depth dependent magnetization in the TmIG bottom layer is simulated in the sample as the exchange length cannot encompass the whole extent of the 40 nm thick TmIG layer. This is clear from the micromagnetic spin configuration indicated in Fig. 8 and the rightmost bottom panel of Fig. 7, which shows the snapshots of the simulated region with different depths in TmIG. The domain wall patterns during the switching process show qualitative matches to that found with MOKE. Domain patterns, at a -3 mT applied field following positive field saturation, exhibit dendritic domains with a maximum width of ~ 1.76 µm (half of the domain wall periodicity), which is qualitatively similar to the width of the domain wall observed in MOKE images.

For $x$ = 4 nm samples, the top layer CoFeB magnetization and the bottom layer TmIG magnetization exhibit smaller projections in the out of plane direction than the x = 1 nm sample as seen from Fig. 7. This is due to the fact that large shape anisotropy forces the CoFeB magnetization to predominantly align along in-plane, and exchange interaction forces the TmIG magnetization into a canted state with the interface magnetization mostly aligned parallel to the CoFeB magnetization (see rightmost top panel of Fig. 7 and Fig. 8a). Thus, magnetostatic interaction mediated by the shape anisotropy governs the coupling between the TmIG and CoFeB magnetization. As a result, weak imprinting of TmIG domains on thick CoFeB is possible, which is predicted by micromagnetic simulation. A similar trend is observed from polar MOKE images where the domains possess a less sharp intensity contrast due to the dominant in-plane magnetization contribution (Fig. 4e). Furthermore, dendritic domains with a maximum size of ~ 0.39 µm are simulated at a 0 mT applied field, which is similar to our MOKE experiment al results (0.46 µm).

The exchange length in CoFeB for the FI/FMM sample with $x$ = 1 nm is, $l_{ex,CoFeB} \approx$ 0.88 nm ($l_{ex} \sim \sqrt{\frac{IEC \cdot c_z}{\frac{1}{2}\mu_0 M_{s,FM}^2}}$, where $c_z$ is the micromagnetic cell size along film normal and $M_{s,FM}$ is the saturation magnetization of the FMM layer). This suggests that most of the CoFeB layer is within the exchange length and coupled strongly with perpendicular TmIG. In FI/FMM with a thicker CoFeB such as $x$ = 4 nm, the exchange is $l_{ex,CoFeB} \approx$ 1.1 nm. This suggests that a large portion of the CoFeB is beyond the exchange length, hence, is not perpendicularly coupled to the TmIG bottom layer. Strong in-plane anisotropy of the CoFeB, in addition to exchange length that is much lower than its thickness, makes the CoFeB magnetizations energetically favorable to be predominantly aligned along the in-plane. Thus, the interaction between the FMM and FI layers is dominated by magnetostatic interaction where the in-plane CoFeB supports flux closure paths during the domain formation in the TmIG layer (Fig. 8a). Although the IEC determined from micromagnetic simulations is higher for heterostructures with thick CoFeB compared to those that are thin, the interlayer exchange lengths in CoFeB are found to be similar for both samples.

This is due to the fact that the saturation magnetization is higher in heterostructure samples with a thick CoFeB layer.

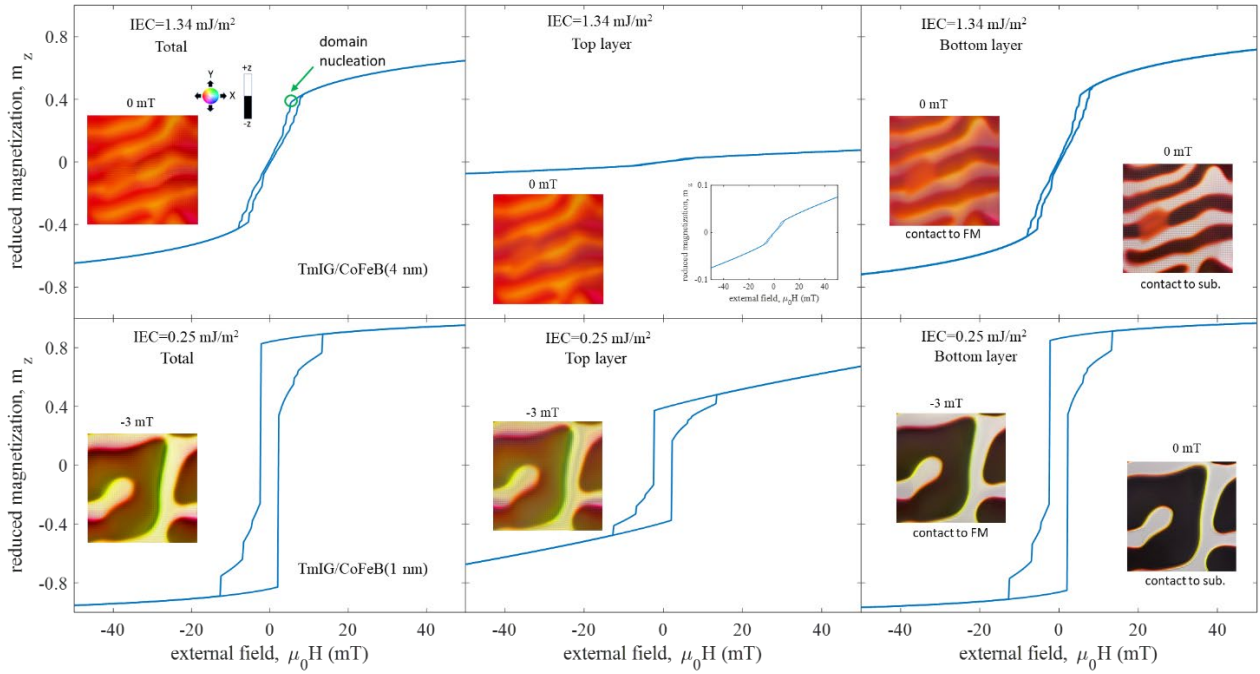

**Figure 7:** Hysteresis loops along the out of plane directions derived from micromagnetic simulations for the total heterostructure, top layer CoFeB and bottom layer TmIG are shown for samples with x=4 nm and x=1 nm. The micromagnetic snapshots for both of the samples for total magnetizations, top CoFeB layer and bottom TmIG layer magnetizations are shown. For the bottom TmIG layer, two snapshots taken at different depths of TmIG (contact to FMM denotes the TmIG layer that is directly in contact with CoFeB and contact to substrate shows the TmIG layer that is in direct contact with the GGG substrate).

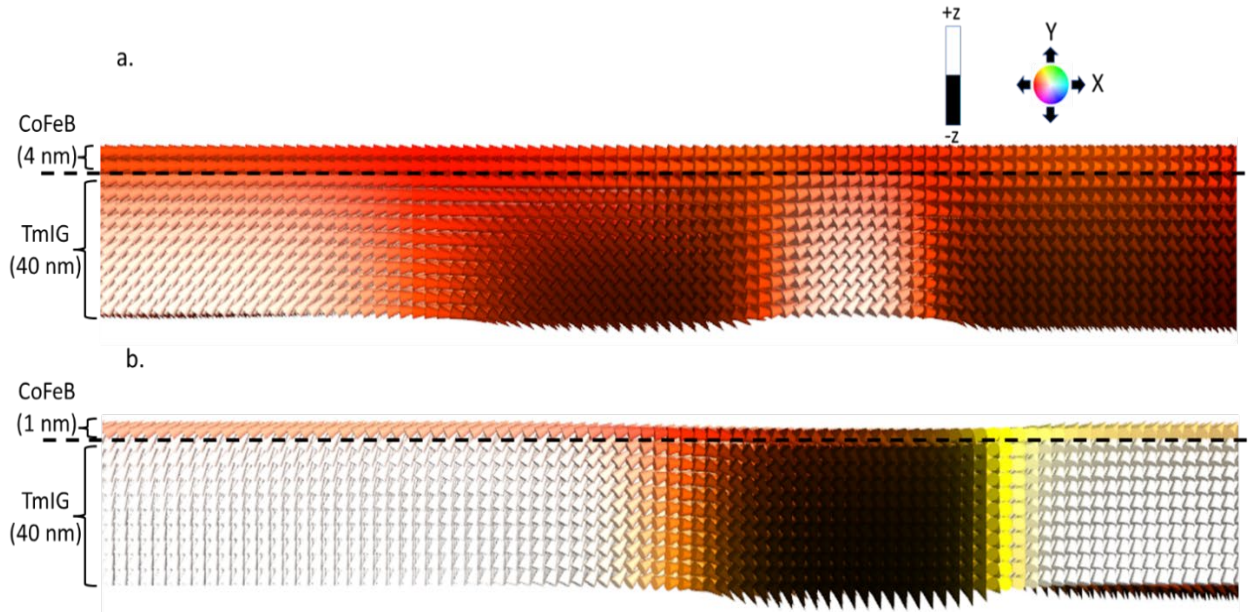

**Figure 8**: Micromagnetic spin orientation in FI/FMM heterostructures with **a.** $x = 4$ nm and **b.** $x = 1$ nm along the cross-section of domain walls for out of plane external field, $\mu_0 H_\perp = 0$ mT and $\mu_0 H_\perp = -3$ mT respectively. The number of spins is down sampled from actual micromagnetic simulations for better visibility.

Micromagnetic analysis is used to determine the coupling strengths using practical material parameters. The out of plane switching process of the FI/FMM heterostructure shows domain wall imprinting of TmIG on CoFeB for samples with thin CoFeB layers (such as $x \leq 1$ nm). The switching in thick CoFeB heterostructure shows a dominant effect from the shape anisotropy of CoFeB which facilitates flux closure.

## VI. Conclusion

The interlayer exchange coupling between a ferrimagnetic insulator with PMA, GGG/TmIG (40 nm) and a ferromagnetic compound CoFeB($x$)/W(0.4 nm)/CoFeB(0.8 nm)/MgO(1 nm)/W(5 nm) is studied by varying the thickness of the variable CoFeB layer. The relative strength of the competing interactions such as the interlayer exchange and magnetostatic coupling is found to be strongly dependent on the thickness of the CoFeB layer. When deposited on silicon, the ferromagnetic stack with CoFeB thickness $x \leq 1$ nm exhibits PMA, however, the PMA is weak and the magnetization switching occurred predominantly in a reversible manner as confirmed by the FORC features. When the same ferromagnetic stack with $x \leq 1$ nm CoFeB is deposited on GGG/TmIG, the PMA of the heterostructure increased compared to the pristine TmIG film. Also, the CoFeB and TmIG magnetizations are observed to switch together with domain nucleation and propagation as evident from the FORC analysis. The dominant exchange coupling is attributed to the modified PMA found in the heterostructures with $x \leq 1$ nm CoFeB. In contrast, magnetostatic coupling dominates the magnetic interaction in heterostructures with an $x \geq 3$ nm CoFeB, as confirmed by the early nucleation of reversed domains from the FORC analysis and the increased number of domain walls indicated by MOKE microscopy. These findings provide important insights into coupling between FI and FMM. Specifically, when the FMM is sufficiently thin ($\leq 1$ nm in our study), the FI domain walls (and in general the magnetization orientation) can be strongly imprinted on the FMM layer. This provide a pathway towards realizing fast and efficient spintronic memory and neuromorphic device control by voltage-induced strain [41-47], current [48-49], and a combination of voltage and current [50-52] where we can take advantage of a magnetostrictive FI with low damping and dissipation and at the same time effectively read out its magnetic state by using exchange coupled FI/FMM as the soft layer of MTJ device.

**ACKNOWLEDGEMENT:**

NSF EECS 1954589 and NSF SHF 1815033. Work at Georgetown University was supported by NSF (ECCS-2151809). The authors acknowledge the use of the Center for Nanoscale Science and Technology (CNST) Nanofabrication facility at the National Institute of Standards and Technology (NIST), Gaithersburg, MD and Nano characterization center (NCC) at VCU. CAR and MJG acknowledge NSF ECCS 2152528 and ECCS 1954606.

# Supplemental Information for

# Interfacial exchange and magnetostatic coupling in a CoFeB/perpendicular ferrimagnetic Thulium Iron Garnet heterostructure

Walid Al Misba [1], Jenae E. Shoup [2], Miela J. Gross [3,4], Dhritiman Bhattacharya [5], Kai Liu [5], Caroline A. Ross [4], Daniel B. Gopman [2], Jayasimha Atulasimha [1,6,7*]

[1] Mechanical and Nuclear Engineering, Virginia Commonwealth University, Richmond, VA, USA
[2] Materials Science & Engineering Division, National Institute of Standards and Technology, Gaithersburg, MD, USA
[3] Electrical Engineering and Computer Science, Massachusetts Institute of Technology, Cambridge, MA, USA
[4] Department of Materials Science and Engineering, Massachusetts Institute of Technology, Cambridge, MA, USA
[5] Department of Physics, Georgetown University, Washington, D.C, USA
[6] Electrical and Computer Engineering, Virginia Commonwealth University, Richmond, VA, USA
[7] Department of Physics, Virginia Commonwealth University, Richmond, VA, USA

*jatulasimha@vcu.edu

## S1. Control layer sequence characterization:

Table SI summarizes the unique layer sequences, along with the corresponding saturation magnetization, easy axis orientation, and the respective coercive field along that orientation. We note for volume estimation, the sample area is calculated using an optical microscope and only the thickness of the magnetic layers are considered.

**Table SI**: Magnetic properties of control layer sequence (step size of applied field is 1 mT and the uncertainty in estimating coercive field is assumed $\pm$ 0.1 mT due to the magnetic field step discretization and fitting error).

| Layer sequence | Saturation magnetization (kA/m) | Easy axis | Coercive field (mT) along easy axis |
|---|---|---|---|
| Si/CoFeB(4 nm)/W(0.4 nm)/CoFeB(0.8 nm)/MgO(1 nm)/W(5 nm) | 970 $\pm$ 7.5 | In-plane | 0.33 |
| Si/CoFeB(3 nm)/W(0.4 nm)/CoFeB(0.8 nm)/MgO(1 nm)/W(5 nm) | 940 $\pm$ 5 | In-plane | 0.25 |
| Si/CoFeB(1 nm)/W(0.4 nm)/CoFeB(0.8 nm)/MgO(1 nm)/W(5 nm) | 545 $\pm$ 9 | Out-of-plane | 0.52 |
| Si/CoFeB(0.84 nm)/W(0.4 nm)/CoFeB(0.8 nm)/MgO(1 nm)/W(5 nm) | 365 $\pm$ 11 | Out-of-plane | 1.54 |

Figure S1 shows the magnetization curves for the Si/CoFeB($x$)/W(0.4 nm)/CoFeB(0.8 nm)/MgO(1 nm)/W(5 nm) samples with $x$ = 0.8 nm, 1 nm, 3 nm and 4 nm. The processed version of these curves (after normalization) is shown in Fig. 2 of the main paper.

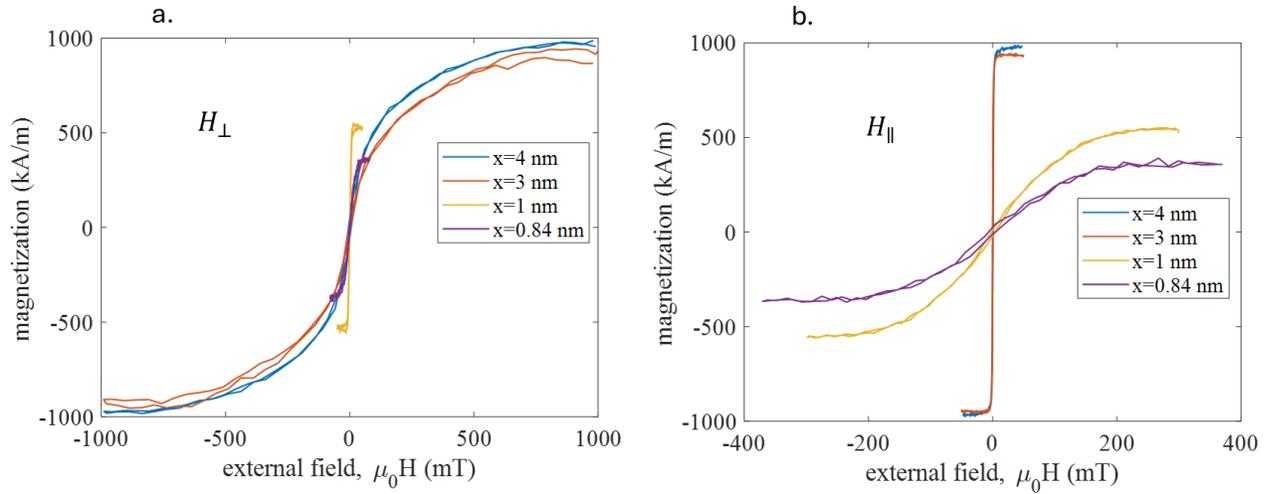

**Figure S1: a.** Out of plane and **b.** in-plane hysteresis loops for control samples, Si/CoFeB(x)/W(0.4 nm)/CoFeB(0.8 nm)/MgO(1 nm)/W(5 nm) for variable thickness, $x$ of CoFeB.

## S2. Ferromagnetic Insulator (FI)-only and Ferromagnetic Insulator (FI)/Ferromagnetic Metal (FMM) Heterostructure Sample Characterization:

The thicknesses of different layers of the heterostructure of the FI-only stack and the FI/FMM stacks along with the measured saturation magnetization, easy axis directions, and the coercive field along the easy axis directions are presented in Table SII. We note for volume estimation, the sample area is calculated using an optical microscope and only the thickness of the magnetic layers are considered.

**Table SII**. Magnetic properties of the FI-only stack and FI/FMM heterostructures (step size of applied field is 1 mT and the uncertainty in estimating coercive field is assumed $\pm$ 0.1 mT due to the magnetic field step discretization and fitting error).

| Layer sequence | Saturation magnetization (kA/m) | Easy axis | Coercive field (mT) along easy axis |
|---|---|---|---|
| GGG/TmIG(40 nm) | 82 $\pm$ 1.6 | Out of plane | 0.22 |
| GGG/TmIG(40 nm)/CoFeB(0.84 nm)/W(0.4 nm)/CoFeB(0.8 nm)/MgO(1 nm)/W(5 nm) | 91.5 $\pm$ 1.7 | Out of plane | 2.03 |
| GGG/TmIG(40 nm)/CoFeB(1 nm)/W(0.4 nm)/CoFeB(0.8 nm)/MgO(1 nm)/W(5 nm) | 95 $\pm$ 1.4 | Out of plane | 1.81 |
| GGG/TmIG(40 nm)/CoFeB(3 nm)/W(0.4 nm)/CoFeB(0.8 nm)/MgO(1 nm)/W(5 nm) | 143 $\pm$ 2.5 | In-plane | 9.60 |
| GGG/TmIG(40 nm)/CoFeB(4 nm)/W(0.4 nm)/CoFeB(0.8 nm)/MgO(1 nm)/W(5 nm) | 169 $\pm$ 2.31 | In-plane | 3.35 |

## S3. Micromagnetic simulation

The simulation geometry is assumed to be 2.048 μm × 2.048 μm × (40+$x$) nm with finite different discrete cells of (4 × 4 × 1) nm$^3$ for FI/FMM heterostructures with $x$ = 4 nm and $x$ = 1 nm. The small lateral dimensions were needed to ensure the micromagnetic simulations are computationally feasible. Furthermore, periodic boundary conditions are used to adequately capture the magnetostatic energy of the larger film geometry. The exchange stiffness of thulium iron garnet (TmIG) is considered to be $A_{FI}$= 2 pJ/m and the uniaxial perpendicular anisotropy is $K_{u,FI}$=7 kJ/m$^3$ consistent with published values [1]. A 5 % variation in $K_{u,FI}$ across grains is considered with an average grain size of approximately 400 nm. The saturation magnetization $M_{s,FI}$ is considered to be 80 kA/m. For the FMM stack, CoFeB($x$)/W(0.4 nm)/CoFeB(0.8 nm)/MgO(1 nm)/W(5 nm), both of the CoFeB layers are considered with a saturation magnetization of $M_{s,FMM}$ at 970 KA/m and 550 kA/m for $x$ = 4 nm and $x$ = 1 nm samples, respectively. For the $x$ = 1 nm sample, the perpendicular anisotropy of MgO/CoFeB is considered to be $K_u$=250 kJ/m$^3$. The exchange stiffness of the CoFeB is considered to be $A_{FI}$= 14 pJ/m [2-3]. The exchange stiffness and saturation magnetization are kept the same for the other CoFeB layer. The exchange interaction between the CoFeB layers separated by a very thin 0.4 nm W spacer is strongly ferromagnetic as seen from the vibrating sample magnetometry (VSM) loops, thus assumed to be only reduced by 20 %. The lateral cell size we considered is within the exchange length of TmIG, $\sqrt{A_{FI}/0.5\,\mu_0 M_{s,FI}^2} = l_{ex} \approx 40$ nm and CoFeB, $\sqrt{A_{FM}/0.5\,\mu_0 M_{s,FMM}^2}$ $l_{ex} \approx 4.9$ nm and $l_{ex} \approx 8.6$ nm for $x$ = 4 nm and $x$ = 1 nm, respectively. A negligibly small external field of the value of, $\mu_0 H_{ext,x}$ =0.001 mT is needed along in-plane x-direction to get convergence while measuring the out of plane hysteresis loops.

**Estimation of optimal interlayer exchange coupling (IEC):**

For the FI/FMM heterostructure with $x$ = 4 nm CoFeB, the sample saturates at $\mu_0 H_\perp$ > 900 mT (not shown in Fig. 7) and the domains starts to nucleate around 5 mT (see green circle in the top panel in Fig. 7), qualitatively similar to what we obtained from VSM measurements where the saturation field is in excess of 900 mT and the domain nucleation starts around 4 mT (when the field is decreased from positive saturation). The pinched magnetic hysteresis loop observed in VSM is well captured in micromagnetic simulation. Increasing the IEC from 1.34 mJ/m$^2$ results in much slower magnetization evolution towards saturation, while decreasing the IEC results in the nucleation field greater than 5 mT. For the FI/FMM sample with $x$ = 1 nm CoFeB, with IEC= 0.25 mJ/m$^2$, the out of plane coercivity predicted from micromagnetic simulation is 2.3 mT which is comparable to the coercivity measured from VSM (1.82 mT). Decreasing the IEC from 0.25 mJ/m$^2$ increases the coercive field from 2.3 mT. Increasing the IEC decreases the coercive field, however in those cases, the magnetization reversal progresses much more slowly compared to what we observed in the VSM loop in Fig. 2a in the main paper.